\documentclass[12pt]{article}

\usepackage{amssymb,amsmath,amsfonts,amssymb}
\usepackage{graphics,graphicx,color,hhline}
\usepackage{bbm} 
\usepackage{authblk}
\usepackage{euscript}
 \usepackage{mathtools}

\def\@abssec#1{\vspace{.05in}\footnotesize \parindent .2in 
{\bf #1. }\ignorespaces} 

\graphicspath{{/EPSF/}{Figures/}}   

\newtheorem{theorem}{Theorem}[section]
\newtheorem{lemma}[theorem]{Lemma}

\def \Rm {\mathbb R}

\def \NN {\mathbb N}

\def \ZZ {\mathbb Z}

\newcommand{\be}{\begin{equation}}
\newcommand{\ee}{\end{equation}}
\newcommand{\bea}{\begin{eqnarray}}
\newcommand{\eea}{\end{eqnarray}}
\newcommand{\bee}{\begin{eqnarray*}}
\newcommand{\eee}{\end{eqnarray*}}

\newcommand{\bk}{\mathbf k}
 \newcommand{\bm}{\mathbf m}

\newcommand{\bu}{\mathbf u} \newcommand{\bv}{\mathbf v}

\newcommand{\bA}{\mathbf A}
\newcommand{\ba}{\mathbf a}

\newcommand{\bx}{\mathbf x} 
\newcommand{\bz}{\mathbf z}

\newcommand{\bK}{\mathbf K}

\newcommand{\bef}{\mathbf e}

\newcommand{\bj}{\mathbf j} 
 
\def\fref#1{{\rm (\ref{#1})}}

\newcommand{\calH}{\mathcal H}

\newcommand{\calV}{\mathcal V}

\newcommand{\calM}{\mathcal M}

\newcommand{\calA}{\mathcal A}
\newcommand{\calD}{\mathcal D}

\newcommand{\bzero}{\mathbf 0}

\newcommand{\cout}[1]{}

\begin{document}
\title{A simple real-space scheme for periodic Dirac operators}
\author{Hua Chen}
{\affil{\small Department of Physics, Colorado State University, Fort Collins, CO 80523, USA}
\affil{School of Advanced Materials Discovery, Colorado State University, Fort Collins, CO 80523, USA}
\author{Olivier Pinaud}
\affil{Department of Mathematics, Colorado State University, Fort Collins, CO 80523, USA}
\author{Muhammad Tahir}
\affil{Department of Physics, Colorado State University, Fort Collins, CO 80523, USA}
\maketitle
\begin{abstract} We address in this work the question of the discretization of two-dimensional periodic Dirac Hamiltonians. Standard finite differences methods on rectangular grids are plagued with the so-called Fermion doubling problem, which creates spurious unphysical modes. The classical way around the difficulty used in the physics community is to work in the Fourier space, with the inconvenience of having to compute the Fourier decomposition of the coefficients in the Hamiltonian and related convolutions. We propose in this work a simple real-space method immune to the Fermion doubling problem and applicable to all two-dimensional periodic lattices. The method is based on spectral differentiation techniques. We apply our numerical scheme to the study of flat bands in graphene subject to periodic magnetic fields and in twisted bilayer graphene.
  \end{abstract}
\section{Introduction}
This work is concerned with the numerical resolution of the stationary Dirac equation with periodic coefficients. The Dirac equation is historically associated with relativistic quantum mechanics, and is central in condensed matter physics in the study of graphene, topological insulators, and Weyl/Dirac semimetals, etc. This is the main application we consider here. The two-dimensional expression of the (linear) Dirac Hamiltonian reads, with all physical constants set to one,
\be \label{Dirac}
H=\sigma \cdot (-i \nabla +A)+\sigma_z M+ I_2 V,
\ee
where $\sigma=(\sigma_x,\sigma_y)$, for $\sigma_j$ the Pauli matrices


$$
\sigma_x=
\begin{pmatrix}
0&1\\
1&0
\end{pmatrix}
\qquad \sigma_y=
\begin{pmatrix}
0&-i\\
i&0
\end{pmatrix}
\qquad \sigma_z=
\begin{pmatrix}
1&0\\
0&-1
\end{pmatrix}.
$$
Above, $I_2$ is the $2\times 2$ identity matrix, $A=(A_1,A_2)$ is the vector potential, $V$ the scalar potential, and $M$ a ``mass'' term. We focus on the 2D case for simplicity of the exposition and the implementation, but the methods we introduce are readily extended to 3D geometries.

In spite of the apparent simplicity of $H$, the discretization of the Hamiltonian is quite subtle. Consider indeed for instance a uniform grid of stepsize $h$ on a square, and discretize the partial derivatives with centered finite differences. When $A$, $M$ and $V$ are all zero, the resulting dispersion relation is, with $\bk=(k_x,k_y)$,
$$
\omega^2(\bk)=\left(\frac{\sin(k_x h)}{h}\right)^2+\left(\frac{\sin(k_y h)}{h}\right)^2, \qquad k_x,k_y \in \left[-\pi h^{-1},\pi h^{-1}\right].
$$
\begin{figure}[h!]
\centering
    \includegraphics[height=6cm, width=8cm]{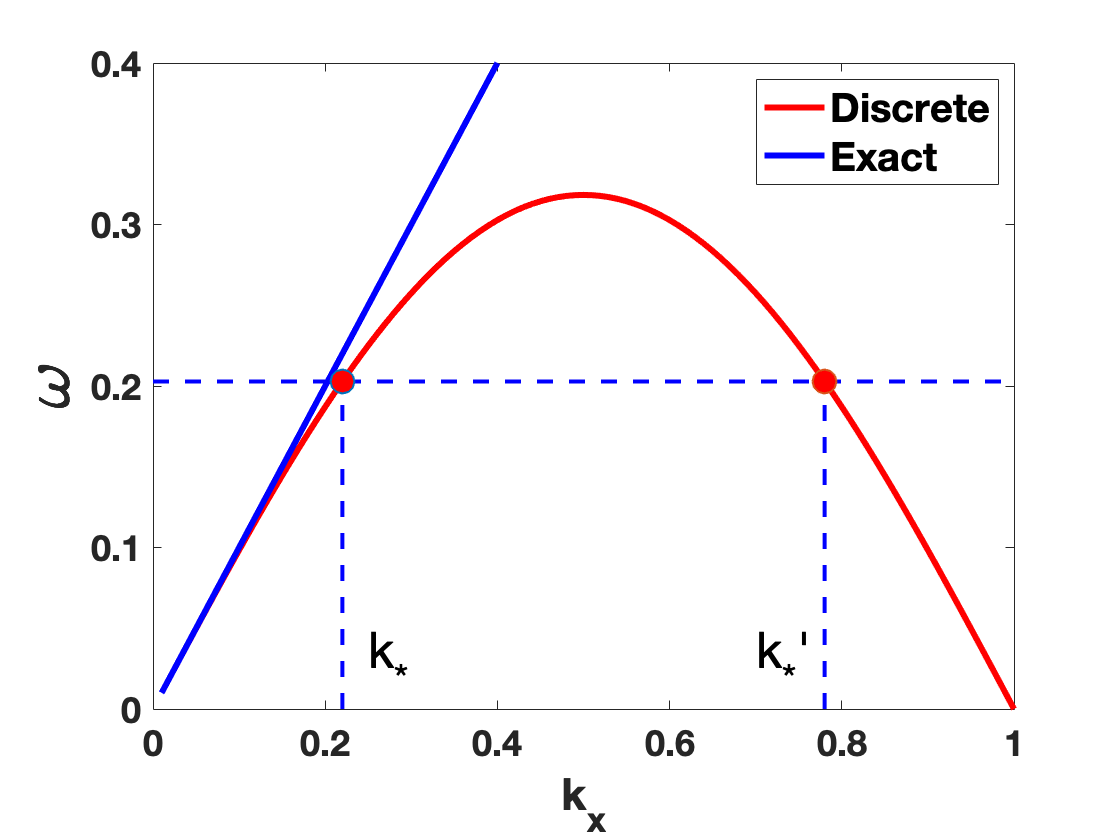} 
\caption{Dispersion relation of the centered finite differences Dirac Hamiltonian.}
    \label{fig:disp}
\end{figure}
It is represented in figure \ref{fig:disp} in the plane $k_y=0$ when $h=\pi$. One recovers the expected relation $\omega(\bk)=|\bk|$ when $|\bk| h \ll 1$. The function $\omega(\bk)$ is not monotone with respect to $k_x$ and $k_y$, and as a consequence there are two solutions to the equation $\omega(\bk)=$constant. The one closest to the origin ($k_\star$ in the figure) is the physical solution approximating the exact solution, while the other one (i.e. $k_\star'$), closer to 1,  is unphysical. It is associated with a highly oscillating mode. Indeed, if the square has side $L$ with periodic boundary conditions, and each direction is discretized with $N+1$ points (with e.g. $N$ even), the eigenvalues of $H$ are
$$
N \sqrt{\left(\frac{\sin(2 \pi m/N)}{L}\right)^2+\left(\frac{\sin(2 \pi n/N)}{L}\right)^2}, \qquad m,n=-N/2,\cdots, N/2,
$$
with eigenvectors $v_{mn}(\ell,\ell')=e^{i 2 \pi (m \ell+n\ell')/N}$, $\ell,\ell'=0,\cdots,N$. Then e.g. $v_{m,n}$ and $v_{N/2-m,n}$ are associated with the same eigenvalue, and the latter is the spurious mode (when $m$ is small). This phenomenon is referred to as the \textit{Fermion doubling problem} in the physics literature, see e.g. \cite{fermion_doubling, fermion_doubling2}. The Schr\"odinger Hamiltonian is immune to the problem as the dispersion relation is monotone, resulting in the eigenvalues
$$
\frac{4N^2}{L^2} \left[\left(\sin(\pi m/N)\right)^2+\left(\sin(\pi n/N)\right)^2\right], \qquad m,n=-N/2,\cdots, N/2.
$$
The large eigenvalues are not properly approximated, but at least spurious modes associated with small eigenvalues are not created. It is then possible to use the ``squaring trick'' to overcome the doubling issue when the coefficients $M$ and $V$ are constant, that is to square the Dirac operator to recover the Schr\"odinger operator. The procedure does not apply when $M$ and $V$ are variable.

A simple way to handle the Fermion doubling problem for linear time evolution problems is to consider initial conditions with only low frequency modes, when possible, that are well propagated by the scheme. The situation is more critical for stationary problems such as band structure calculations where the discrete Hamiltonian is diagonalized. In this case, the dimension of the eigenspaces is wrongly doubled and, as a consequence, while the low eigenvalues that are calculated are accurate approximations, some physical ones are left out by the procedure. Besides, the eigenvectors are corrupted since any linear combination of the physical eigenvector and of the spurious one is also an eigenvector. 

The classical method for band structure calculations found in the physics community consists of using so-called ``plane-wave expansions''. The idea is simply to perform a Fourier transform adapted to the lattice, resulting in an exact dispersion relation. The major disadvantage is that the Fourier coefficients of the functions $A$, $M$, $V$ have to be computed along with some convolutions. It is then customary to find in the literature simple coefficients $A$, $M$, $V$ with just a small number of Fourier modes that are known by construction. This limits the applicability of the method as one would like to use arbitrary coefficients in the Dirac Hamiltonian.

It is then desirable to derive methods in real-space. One-sided derivatives that are widely used in the context of first-order hyperbolic equations such as the Dirac equation are not appropriate since they break the hermicity of the Dirac Hamiltonian. Some solutions were proposed in \cite{anton_dirac3D,anton1D} in the time-dependent case, and could be adapted to the periodic stationary picture. They have two main limitations: they are designed for cartesian grids, which is a serious impediment for band structure calculations, and they are quite technically involved. Indeed, the number of unknowns is doubled in these methods, and they are defined on two different staggered grids resulting in a somewhat complicated scheme.

We propose in this work a simple real-space scheme overcoming the Fermion doubling problem and applicable to all two-dimensional periodic lattices. The scheme is based on classical spectral methods, see e.g. the monograph \cite{T-Spectral-00}. These methods rely on Fourier series expansions, and are therefore capturing the dispersion relation exactly and are perfectly adapted to the periodic setting. These methods essentially ``bring back'' the plane-wave expansion methods of the physics community to the real space. The Hamiltonian is discretized in the primitive cell of the lattice, which takes the form of a parallelogram, see e.g. the Honeycomb (i.e. hexagonal) and Kagome lattices in figure \ref{fig:lat}.
\begin{figure}[h!]
\centering
\vspace{-3.8cm}
    \includegraphics[height=12cm, width=9cm]{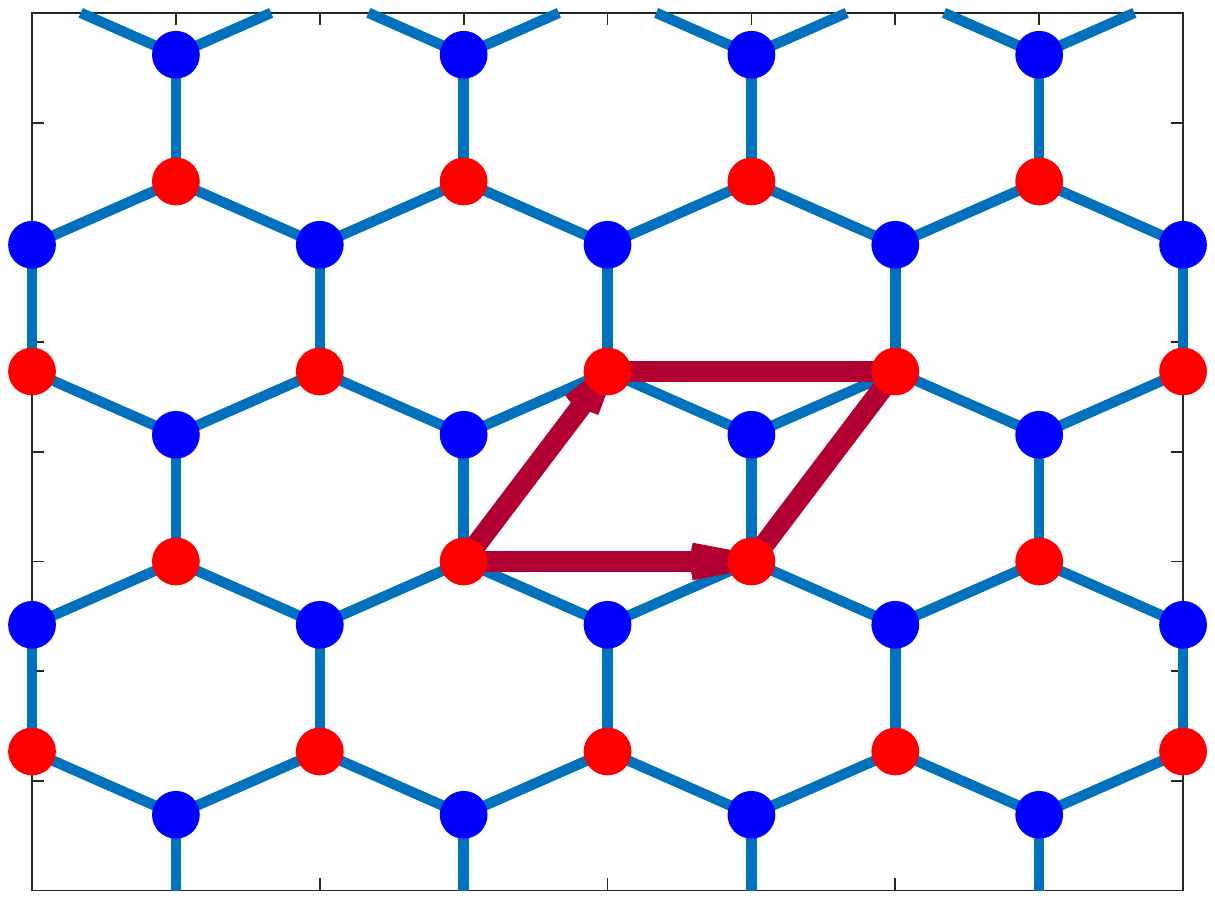} \hspace{-3cm}
    \includegraphics[height=12cm, width=9cm]{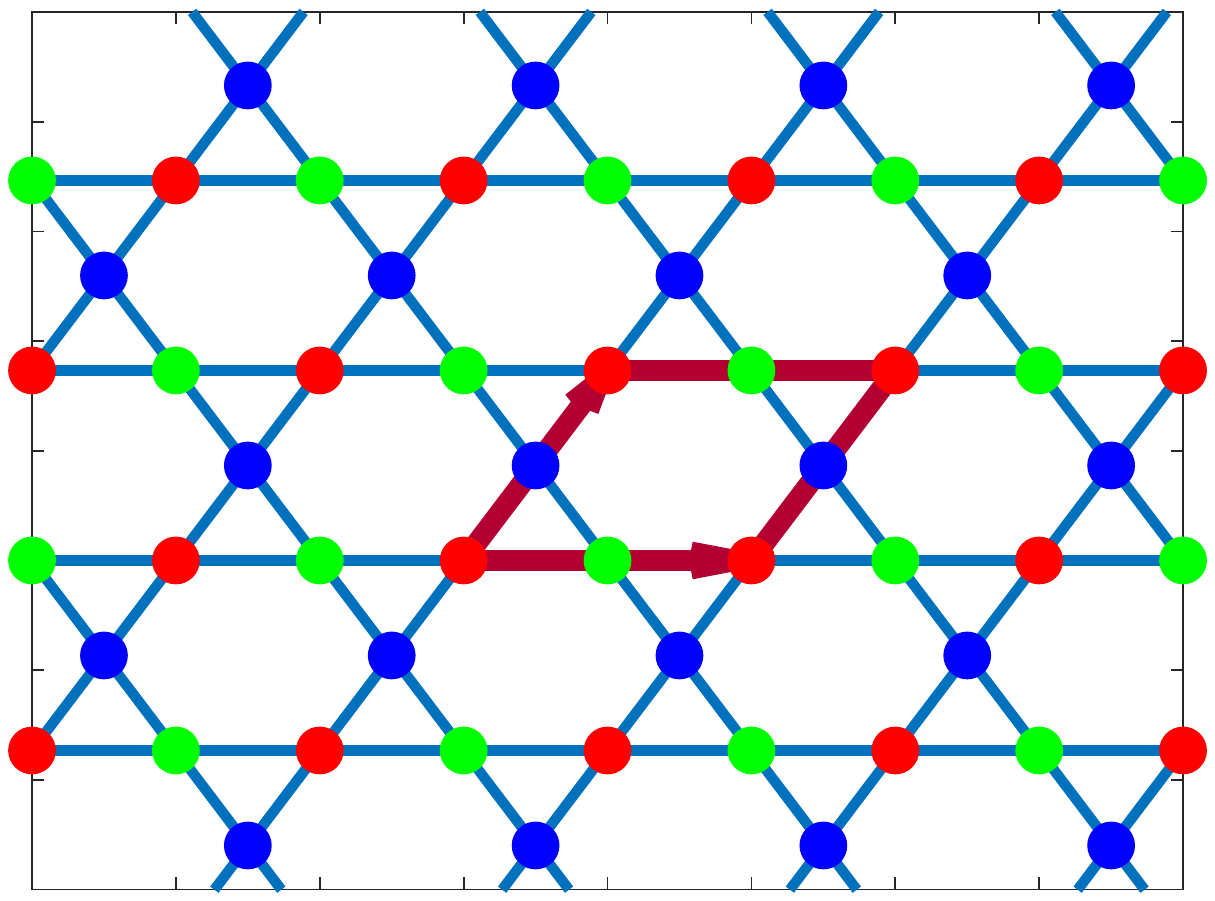}
\vspace{-3.8cm}
\caption{Left: Hexagonal lattice; Right: Kagome lattice. The primitive cell is depicted in red along with the primitive vectors.}
    \label{fig:lat}
\end{figure}
The partial derivatives are replaced by so-called differentiation matrices, and rotations are performed to align the directions of derivation with the primitive vectors of the lattice. One has to be careful with the construction of these differentiation matrices since, in the standard setting where the number of discretization points is even, the kernels of the matrices are two-dimensional and spanned by highly oscillating sawtooth functions, instead of simply consisting of the constant vectors. This is not a major issue for evolution problems, but is a critical one for eigenvalue calculations as any eigenvector associated with a nonzero eigenvalue $\lambda$ can be multiplied component-wise by an eigenvector in the kernel and still be associated with $\lambda$. The eigenvectors can then be corrupted by eigenvectors in the kernel and exhibit spurious unphysical oscillations. A simple way around this is to consider an \textit{odd} number of discretization points, resulting in differentiation matrices with one-dimensional kernels spanned by constant vectors.

We apply our numerical method to the study of flat bands in graphene. These have received much interest lately as they tend to promote interaction-driven ordering phenomena such as unconventional superconductivity since the kinetic energy of the particles is small. We consider two situations. The first one is graphene in a periodic magnetic field. The second one is twisted bilayer graphene consisting of two graphene sheets on top of each other and rotated by a given angle. The resulting Hamiltonian consists of two coupled Dirac Hamiltonians, and flat bands are observed for a particular set of angles. We will study numerically the stability of these flat bands with respect to random perturbations. Another potential application of our method that is not considered here, is the setting of \cite{uri}, where the twisting angle is position-dependent due experimental uncertainties across the structure.

The article is structured as follows: we describe our numerical method in section \ref{meth}, and section \ref{appli} is devoted to the applications.

\paragraph{Acknowledgement.} OP is supported by NSF CAREER grant DMS-1452349. HC and MT are supported by the start-up funding from CSU.

%

\section{The numerical method} \label{meth}
We start by setting the geometry.
\subsection{Preliminaries}

We use a particular set of coordinates to derive the scheme, but other choices are possible. We suppose a rotation has been performed beforehand so that one of the primitive vectors is aligned with the $x$ axis. We denote by $\Lambda$ the primitive cell of the lattice, with the lower left node located by convention at $(0,0)$. We denote by $\ba_1$ and $\ba_2$ (not normalized to one) the vectors generating the lattice, and by $\bk_1$, $\bk_2$ the ones generating the reciprocal lattice. They are related by
$$
\bk_i \cdot \ba_j= 2 \pi \delta_{ij}, \qquad i,j=1,2.
$$
The vectors $\bk_1$ and $\bk_2$ admit the expressions
$$
\bk_1=\frac{2 \pi R \ba_2}{\ba_1 \cdot R \ba_2}, \qquad \bk_2=\frac{2 \pi R \ba_1}{\ba_2 \cdot R \ba_1},
$$
where $R$ denotes 90 degrees rotation. The cartesian unit vectors associated with the $x$ and $y$ axes are $\bef_1$ and $\bef_2$, see figure \ref{fig:ax}.
\begin{figure}[h!]
\centering
\vspace{-4.3cm}
    \includegraphics[height=14cm, width=12cm]{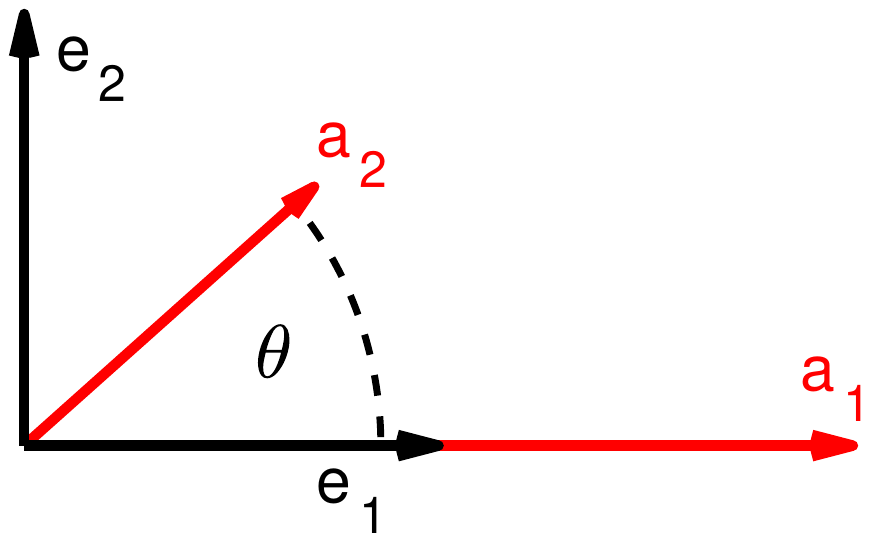} 
\vspace{-5.9cm}
\caption{Axes}
    \label{fig:ax}
  \end{figure}
Let $a_2=|\ba_2|$ and $a_2=|\ba_2|$ be the lengths of $\ba_1$ and $\ba_2$, and let $\bu_1=\ba_1/a_1$, and $\bu_2=\ba_2/a_2$ be unit vectors. A point in the primitive cell is denoted by $\bu=u_1 \bu_1+u_2 \bu_2$, with $0\leq u_{1}\leq a_{1}$ and $0\leq u_{2}\leq a_{2}$. If $\theta \neq 0$ is the angle between $\ba_1$ and $\ba_2$, the cartesian coordinates $(x_1,x_2)$ of $\bu=x_1 \bef_1+x_2 \bef_2$ are related to $(u_1,u_2)$ by 
$$
x_1=u_1+u_2 \cos (\theta), \qquad x_2=u_2 \sin(\theta),
$$
or equivalently
\be \label{CV}
u_2=x_2/\sin(\theta), \qquad u_1=x_1-x_2 \cot(\theta).
\ee
We get in particular the following relations between the partial derivatives that will be used further:
$$
\partial_{x_1}=\partial_{u_1}+\cos(\theta) \partial_{u_2}, \qquad \partial_{x_2}=\sin(\theta) \partial_{u_2}.
$$

We now turn to the approximation of the Dirac Hamiltonian.

\subsection{Discretization}  Let $f$ be a periodic function over the lattice such that $\|f\|^2_{L^2(\Lambda)}=\int_{\Lambda} |f(\bu)|^2 d\bu$ is finite. Then $f$ can be decomposed into the Fourier series
$$
f(\bu)=\sum_{\bm \in \ZZ^2} e^{i \bK_\bm \cdot \bu} \hat f_\bm,
$$
with convergence in $L^2(\Lambda)$. Above, $\bK_\bm=m_1 \bk_1+m_2 \bk_2$ with $\bm=(m_1,m_2) \in \ZZ^2$, and the $\hat f_\bm$ are the Fourier coefficients of $f$. In particular, the delta function reads in the distribution sense
$$
\delta(\bu)=\sum_{\bm \in \ZZ^2} e^{i \bK_\bm \cdot \bu}.
$$
Consider now the following discretization of $\Lambda$: for $Q_1$ and $Q_2$ given in $\NN$, let $\bj=(j_1,j_2)$, with $j_1=1,\cdots N_1$, $j_2=1,\cdots N_2$, and $N_1=2Q_1+1$, $N_2=2Q_2+1$; a grid point in $\Lambda$ is then denoted by $\bu_{\bj}=j_1 h_1 \bu_1+j_2 h_2 \bu_2$, for $h_1=a_1/N_1$, $h_2=a_2/N_2$. With such a choice for $\bj$, the bottom and left sides of the primitive cell are ignored in the discretization because of the periodicity of the lattice.

The next result is standard: we have
$$\frac{1}{N_1}\sum_{|m_1|\leq Q_1}e^{\frac{i 2 \pi m_1 u_1}{a_1}} = D_{Q_1}\left(\frac{2 \pi  u_1}{a_1}\right), \qquad \textrm{where} \qquad D_n(x)=\frac{\sin\big((n+1/2)x\big)}{(2n+1) \sin(x/2)}.
$$
The function $D_n$ is commonly referred to as the Dirichlet kernel (up to the normalization factor). The series defining the delta function is then truncated and $\delta$ is approximated by 
$$
\delta_a(\bu)=\frac{1}{N_1N_2}\sum_{|m_1|\leq Q_1}\sum_{|m_2| \leq Q_2} e^{\frac{i 2 \pi m_1 u_1}{a_1}} e^{\frac{i 2 \pi m_2 u_2}{a_2}}=D_{Q_1}\left(\frac{2 \pi  u_1}{a_1}\right) D_{Q_2}\left(\frac{2 \pi  u_2}{a_2}\right).
$$
The normalization is chosen such that $\delta_a(\bzero)=1$, with $\bzero=(0,0)$. We introduce the following notations for simplicity
$$
S_1(x)=D_{Q_1}\left(\frac{2 \pi  x}{a_1}\right), \qquad
S_2(x)=D_{Q_2}\left(\frac{2 \pi  x}{a_2}\right).
$$
Owing to the relation
$$
f_{\bj}:=f(\bu_{\bj})=\sum_{\bm \in \ZZ^2} f_{\bm} \delta(\bj-\bm),
$$
the function $f$ is then approximated at $\bu \in \Lambda$ by
\bea \label{appf}
f_a(\bu)&=& \sum_{|m_1|\leq Q_1}\sum_{|m_2| \leq Q_2} f_{a,\bm} \delta_a(\bu-m_1 h_1 \bu_1-m_2 h_2 \bu_2) \nonumber\\
&=&\sum_{|m_1|\leq Q_1}\sum_{|m_2|\leq Q_2} f_{a,\bm} S_{1}(u_1-m_1 h_1)S_{2}(u_2-m_2 h_2),
\eea
with $f_{a,\bm}:=f_a(u_\bm)$. The coefficients $f_{a,\bm}$ are defined by periodicity for nonpositive indices $m_1$ and $m_2$: with the notation $f_{a,\bm}=f_{a, m_1,m_2}$, we have e.g. $f_{a,\bm}=f_{a,m_1,m_2}=f_{a,Q_1-m_1,m_2}$ when $-Q_1 \leq m_1 \leq 0$. We will drop in the sequel the index $a$ with an abuse of notation to lighten the expressions. Hence, $f_{a,\bm}$ will become $f_\bm$.

\paragraph{Derivatives.} With \fref{appf}, it is then direct to compute the partial derivatives of the approximate function $f_a$. Indeed, with \fref{CV}, we find
\bee
\partial_{x_1} f_a(\bu)&=&\sum_{|m_1|\leq Q_1}\sum_{|m_2|\leq Q_2} f_{\bm} S'_{1}(u_1-m_1 h_1)S_{2}(u_2-m_2 h_2)\\
\partial_{x_2} f_a(\bu)&=&- \cot(\theta)\sum_{|m_1|\leq Q_1}\sum_{|m_2|\leq Q_2} f_{\bm} S'_{1}(u_1-m_1 h_1)S_{2}(u_2-m_2 h_2)\\
&&+ \frac{1}{\sin(\theta)}\sum_{|m_1|\leq Q_1}\sum_{|m_2|\leq Q_2} f_{\bm} S_{1}(u_1-m_1 h_1)S'_{2}(u_2-m_2 h_2).
\eee
At one of the grid points $\bu=\bu_\bj$, we have
$$
S'_{1}(j_1 h_1)=\frac{\pi}{a_1} \frac{(-1)^{j_1}}{\sin(j_1 \pi/N_1)}, \qquad S'_{2}(j_2 h_2)=\frac{\pi}{a_2} \frac{(-1)^{j_2}}{\sin(j_2 \pi/N_2)}, \qquad S'_1(0)=S_2'(0)=0,
$$
leading to the expressions, for $j_1=-Q_1,\cdots, Q_1$ and $j_2=-Q_2,\cdots, Q_2$,
\bee
\partial_{x_1} f_a(\bu_\bj)&=&\sum_{|m_1|\leq Q_1}f_{m_1,j_2} S'_{1}\big((j_1-m_1)h_1\big)\\
\partial_{x_2} f_a(\bu_\bj)&=&- \cot(\theta)\sum_{|m_1|\leq Q_1}f_{m_1,j_2} S'_{1}\big((j_1-m_1)h_1\big)\\
&&+\frac{1}{\sin(\theta)}\sum_{|m_2|\leq Q_2} f_{j_1,m_2} S'_{2}\big((j_2-m_2)h_2\big).
\eee
As before, $\partial_{x_1} f_a(\bu_\bj)$ is extended by periodicity: e.g. $\partial_{x_1} f_a(\bu_{-j_1,j_2})=\partial_{x_1} f_a(\bu_{N_1-j_1,j_2})$ for $j_1=0,\cdots,Q_1$. In order to write the derivatives in a compact form, the $N_1 \times N_2$ matrix $f_{\bj}$ is stored into a vector $F$ of length $N_1N_2$, with the correspondence $F_{j_1+N_1(j_2-1)}=f_{j_1,j_2}$ for $j_1=1,\cdots,N_1$, $j_2=1,\cdots,N_2$. Introducing the antisymmetric matrices $T_1$ and $T_2$ defined by $(T_1)_{ij}=S_1'\big((i-j) h_1\big)$ and $(T_2)_{ij}=S_2'\big((i-j) h_2\big)$, $T_1$ of size $N_1 \times N_1$ and $T_2$ of size $N_2 \times N_2$, the partial derivatives of a function $f$ at $\bu_\bj \in \Lambda$ are then approximated by
$$
\partial_{x_1} f(\bu) \longrightarrow \calD_1 F, \qquad \partial_{x_2} f(\bu) \longrightarrow \calD_2 F,
$$
where $$
\calD_1= I_{N_2}  \otimes T_1 , \qquad \calD_2=-\cot(\theta)\; I_{N_2}  \otimes T_1+\frac{1}{\sin(\theta)} \; T_2 \otimes I_{N_1}.
$$
Above, $I_N$ is the identity matrix of size $N \times N$ and $\otimes$ the tensor product. The matrices $T_j$ are full, while the $\calD_j$ are sparse with $N_1 N_2 N_j$ non-zero elements. How good an approximation the differentiation matrices $T_1$ and $T_2$ provide us with depends on the regularity of $f$. If $f$ is for instance analytic, then there is spectral convergence, i.e. the error decreases exponentially with $N_1$ and $N_2$, see e.g. \cite{T-Spectral-00}, Chapters 1 and 4.

The next Lemma shows that the kernel of the differentiation matrix $T_1$ for $N_1$ odd is one-dimensional and spanned by constant vectors. As already mentioned in the introduction, this is to be contrasted with differentiation matrices for $N_1$ and $N_2$ even that have two-dimensional kernels spanned by the highly oscillating sawtooth function, see \cite{T-Spectral-00}, Chapter 3. 
\begin{lemma} \label{kernel} When $N_1$ (resp. $N_2$) is odd, the kernel of the matrix $T_1$ (resp. $T_2$) consists of constant vectors  of length $N_1$ (resp. $N_2$) of the form $C (1, \cdots, 1)$ for $C \in \Rm$.
  \end{lemma}

The proof of the Lemma is elementary and based on the discrete Fourier transform. It is given in Appendix for the reader's convenience. From Lemma \ref{kernel}, it follows that the kernels of $\calD_1$ and $\calD_2$ are one-dimensional and spanned by the constant vector.

\paragraph{The Dirac Hamiltonian.} Using the calculations of the previous section, the partial derivatives in the Dirac operator $H$ defined in \fref{Dirac} are replaced by the appropriate differentiation matrices, and the  discrete version of $H$ is
$$
\calH= \sigma_1 [-i \calD_1+\calA_1]+ \sigma_2 [-i \calD_{2}+\calA_2]+\sigma_3 [\calM]+\calV I_{2N},
$$
where, for $N=N_1N_2$, $\bzero_{N}$ the zero matrix of size $N \times N$, and $B$ any $N \times N$ matrix,
$$
\sigma_1[B]=
\begin{pmatrix}
\bzero_{N}&B\\
B&\bzero_{N}
\end{pmatrix}
\qquad \sigma_2[B]=i
\begin{pmatrix}
\bzero_{N}&-B\\
B&\bzero_{N}
\end{pmatrix}
\qquad \sigma_3[B]=
\begin{pmatrix}
B&\bzero_{N}\\
\bzero_{N}&-B
\end{pmatrix},
$$
and where $\calA_1$, $\calA_2$, $\calM$$, \calV$ are diagonal matrices with diagonals given by the values of the coefficients $A_1$, $A_2$, $M$, $V$ at the grid points $\bu_\bj$ and stored as explained in the previous paragraph. The $2N \times 2N$ matrix $\calH$ is hermitian by construction.
 
With $\calD_\pm=-i\calD_1\pm\calD_2$, $\calA=\calA_1+i\calA_2$, $\calH$ is recast as
$$
\calH=
\begin{pmatrix}
\calV+\calM&\calD_-+\calA^*\\
\calD_++\calA&\calV-\calM
\end{pmatrix}.
$$

\paragraph{The band structure.} With $\bu=x_1 \bef_1+x_2 \bef_2$,  the band structure is computed by considering the family of Hamiltonians $H(\bk):= e^{-i \bk \cdot \bu} \circ H \circ e^{i \bk \cdot \bu} $ acting on periodic functions on the lattice. The vector $\bk$ is defined by $\bk=k_1 \bef_1+k_2 \bef_2$ and belongs to the first Brillouin zone of the reciprocical lattice. The operator $H(\bk)$ is simply obtained by shifting $A_1$ and $A_2$ by $ k_1$ and $ k_2$, respectively. The corresponding discrete operator is then 
$$
\calH(\bk)= \sigma_1 [-i \calD_1+\calA_1+k_1 I_{N}]+ \sigma_2 [-i \calD_{2}+\calA_2+k_2I_{N}]+\sigma_3 [\calM]+\calV I_{2N}.
$$
With $\bK=(k_1+ik_2) I_{N}$, this can be recast as
$$
\calH(\bk)=
\begin{pmatrix}
\calV+\calM&\calD_-+\calA^*+\bK^*\\
\calD_++\calA+\bK&\calV-\calM
\end{pmatrix}.
$$
The band structure then follows by diagonalizing $\calH(\bk)$ for each $\bk$.

\section{Applications} \label{appli}
We apply now the methods of the previous section to the study of flat bands in graphene. The first application is graphene in periodic magnetic fields. 
\subsection{Asymptotic flat bands in periodic magnetic fields}
We follow the setting of \cite{TPC-bands}. Including the physical constants and setting $M=0$ and $V=0$, the Hamiltonian \fref{Dirac} becomes
$$
H=v_F \sigma \cdot \Pi, \qquad \Pi=-i\hbar \nabla+e \bA,
$$
where $v_F$ is the Fermi velocity, $e$ the absolute value of the electron charge, and $\bA$ the vector potential periodic on a square lattice. We consider two forms for $\bA$. The first one is a simple sinusoidal potential that reads
$$
\bA(\bx)=\frac{B}{K} \left(-\sin(K x_2),\sin(K x_1)\right),
$$
where $B$ is the strength of the magnetic field and $K$ the wavenumber. The period is then $\lambda=2 \pi/K$. The associated magnetic field has zero average on each cell in the lattice, and we chose here to center the primitive cell at the origin. Nondimensionalizing and keeping the same notations, the primitive cell is $\Lambda=[-\pi,\pi] \times [-\pi,\pi]$, and we find
$$
H=\sigma \cdot \Pi, \qquad \Pi=-i  \nabla+\bA, 
$$
with
$$
\bA(\bx)=t (-\sin(x_2) \bef_1+\sin(x_1) \bef_2), \qquad t=\frac{eB}{\hbar K^2}.
$$
The second form of $\bA$ is more realistic and suggested in \cite{zhai}, equation 11, and is accompanied by a scalar field $V_s$. We have, for $\bx$ in the same cell $\Lambda$ as above,
\bee
\bA_s(\bx)&=&-\tau g\left(\frac{r}{\sigma}\right) \left(\cos(2\theta) \bef_1-\sin(2\theta) \bef_2\right)\\
V_s(\bx)&=& \eta \tau g\left(\frac{r}{\sigma}\right).
\eee
Above, $g(r)=r^2 e^{-r^2}$, and $(r,\theta)$ are the polar coordinates $x_1=r\cos(\theta)$, $x_2=r\sin(\theta)$. These potentials are strain fields induced by the buckling of the structure, and it is not difficult to show that the associated pseudo magnetic field has zero average in $\Lambda$. The fields $\bA_s$ and $V_s$ are not periodic as defined, but assuming that $\sigma$ is sufficiently small, they are close to zero at the edges of the primitive cell $\Lambda$ and can be assumed to be periodic. The parameters $\tau$ and $\eta$ measure the strength of the strain fields, and $\tau$ is proportional to the period $\lambda$ because of the nondimensionalization. The new Hamiltonian is then
$$
H_s=\sigma \cdot (-i  \nabla+\bA_s)+I_2 V_s.
$$

We compute now the band structure of $H$ and $H_s$ using the method of the previous section. Recall that the ``squaring trick'' does not apply here because of the variable potential $V_s$, and we have therefore to work with the Dirac Hamiltonian. Moreover, even though the Fourier coefficients of $\bA$ are trivial to obtain, one would have to compute those of $\bA_s$ and $V_s$ in order to use the plane-wave expansion method, while this is not needed for our method. We set $N_1=N_2=25$, resulting in a $1250 \times 1250$ sparse matrix $\calH(\bk)$ with at most $2N_1N_2+2N_1N_2(N_1+N_2)=63750$ nonzero elements. We use MATLAB to implement the scheme, and the function \texttt{eigs} to compute only a small number of eigenvalues of interest. 

In figure \ref{fig:VeffM}, we represent the band structure around the zero energy for the sinusoidal field along the $k_x=k_1$ direction in the Brillouin zone. The Dirac point is located at $\bk=(0,0)$. The flattening of bands around the zero energy is clearly observed as $t$ increases. What makes this behavior interesting from a physical viewpoint is the fact that the magnetic field has zero average. It is indeed well-known that strong uniform magnetic fields tend to confine particles, while here the localization has an additional contribution from the $\pi$ Berry phase of Dirac electrons, especially for small magnetic fields (this leads to the Zeeman potential in the squared Hamiltonian which localizes the wavefunctions for small magnetic fields). The nondimensional parameter $t$ can be increased by increasing the strength $B$ of the magnetic field, but also and more interestingly by decreasing the wavenumber $K$. There is a physical lower bound for $K$ set by the disorder potential in real materials, see \cite{TPC-bands} for more details, and therefore a limit on how flat the band can be. This asymptotic flatness is to be contrasted with the one of Moir\'e structures discussed in the next section, which requires fine tuning for flat bands to appear while here flatness is obtained by increasing $t$ monotonically.

\begin{figure}[h!]
\centering
\includegraphics[height=5.5cm, width=7cm]{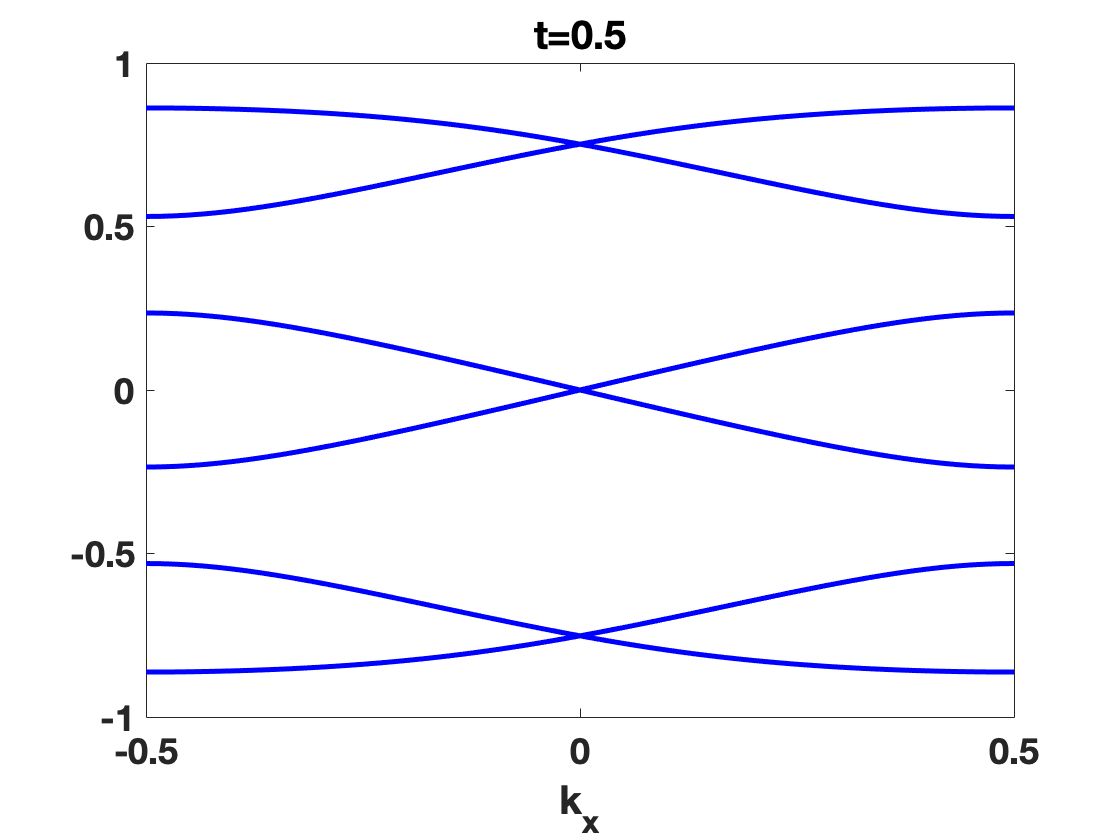}
    \includegraphics[height=5.5cm, width=7cm]{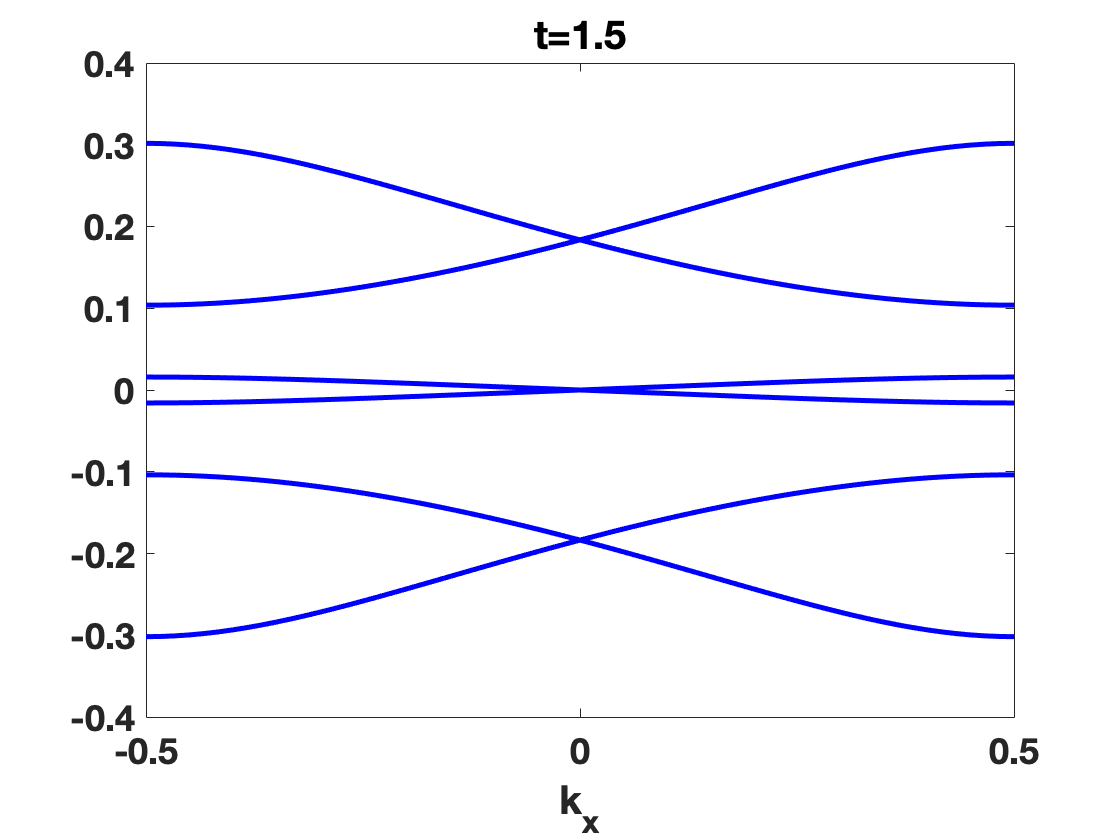} 

\caption{Band structure around the zero energy for the sinusoidal field. The Dirac cone around the origin flattens as $t$ increases.}
    \label{fig:VeffM}
  \end{figure}

We quantify the flatness in figure \ref{fig:VeffM2} and represent the effective velocity for both the sinusoidal and gaussian fields. If the dispersion relation around the origin is $\omega(\bk)=\alpha |\bk|$ for the nondimensional problem, then the effective velocity is defined by $v^{\rm{eff}}_F=\alpha v_F$. The left panel corresponds to the sinusoidal case, and exhibits the monotonic decrease of the effective velocity as $t$ increases. On the right panel, we represent $v^{\rm{eff}}_F$ when $\eta=0$ (full lines) and $\eta=0.05$ (dashed lines). We only consider small scalar fields as stronger ones would change the location of the Dirac point. The width refers to the parameter $\sigma$ in the definition of the fields, and the period is $2 \pi$. We observe the same monotone behavior as the strength (here $\tau$) increases. In particular, this suggests that asymptotic flatness does not depend on the particular form of the magnetic field. We refer to \cite{TPC-bands} for more comments about flat bands in periodic magnetic fields.

\begin{figure}[h!]
\centering
\includegraphics[height=5.5cm, width=7cm]{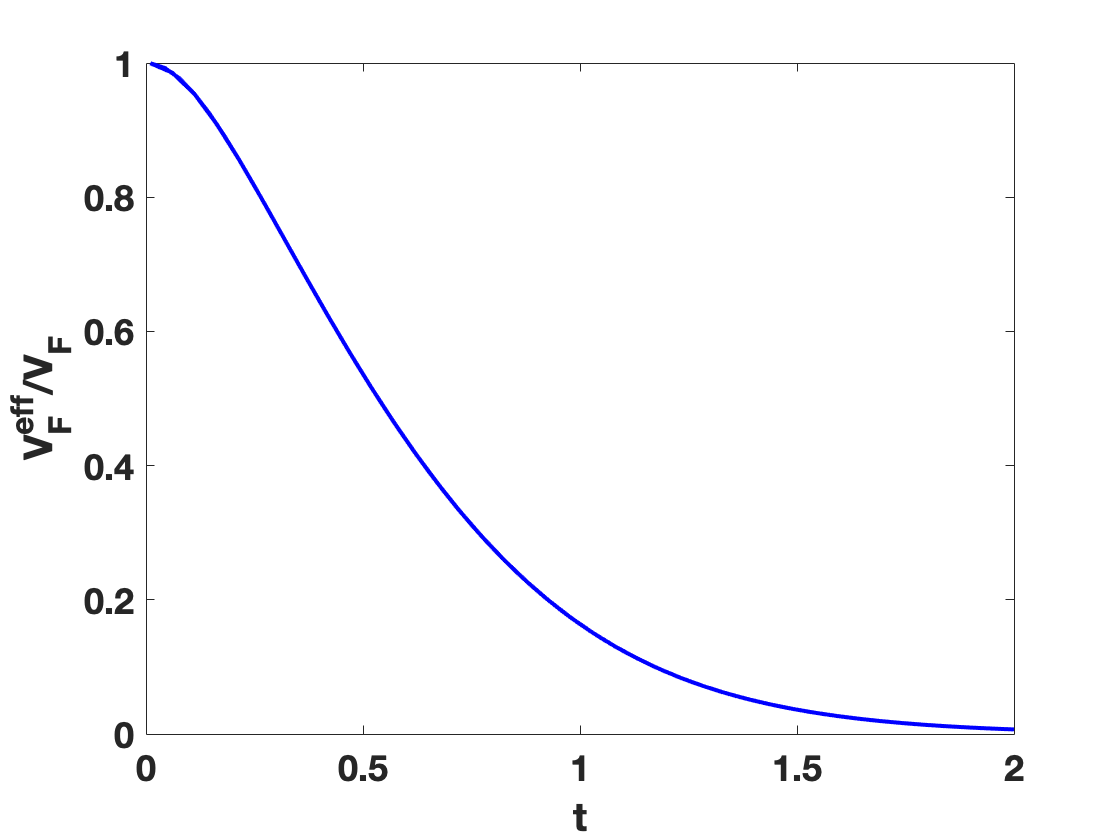}
    \includegraphics[height=5.5cm, width=7cm]{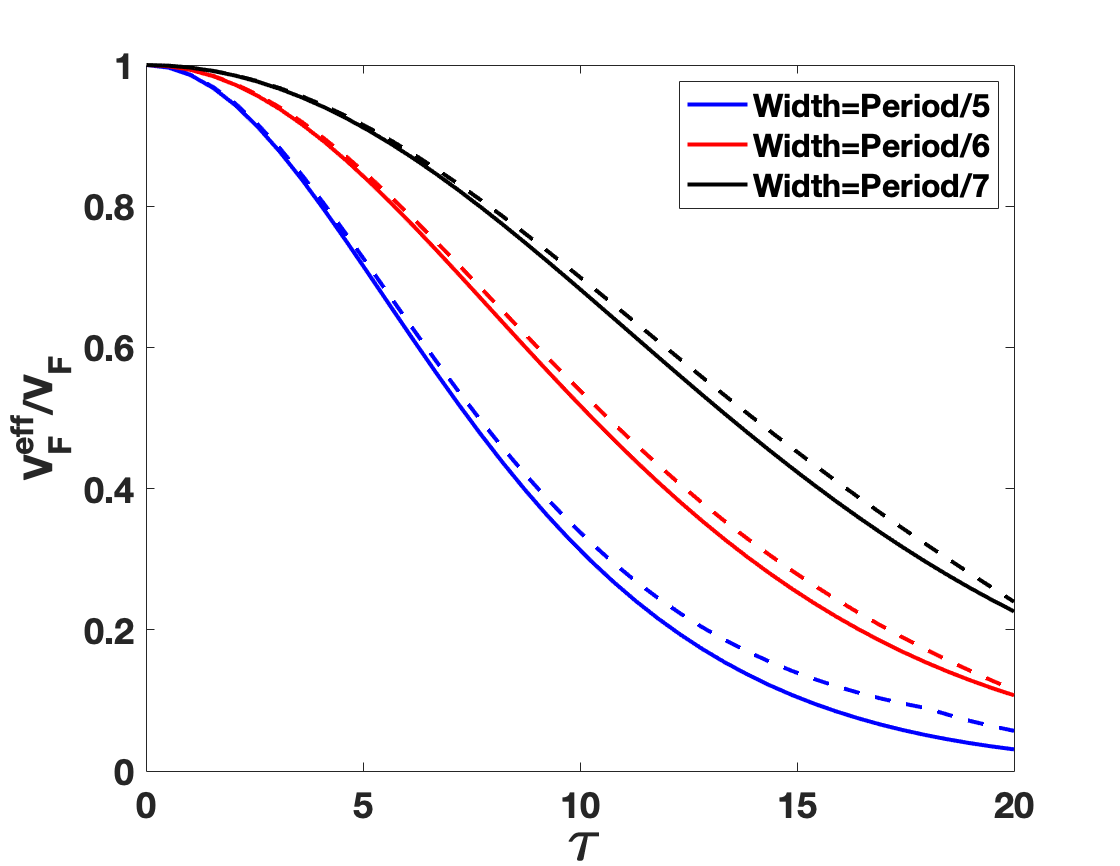} 

\caption{Effective velocity as a function of the field strength. Left: sinusoidal field. Right: gaussian field, with $\eta=0$ (full lines) and $\eta=0.05$ (dashed lines). A similar monotone behavior is observed in both cases.}
    \label{fig:VeffM2}
  \end{figure}

  We turn now to our second application concerned with twisted bilayer graphene.
  
  \subsection{Flat bands in twisted bilayer graphene}
  The model simulated in this section is based on \cite{sanjose}. It consists of two graphene sheets on top of each other and twisted by a tunable angle. The twisting creates Moir\'e patterns, and it has been observed that for a set of so-called ``magic angles'', the zero energy band is essentially flat, leading to interesting physical phenomena. In particular, unconventional superconductivity has been observed experimentally at the magic angles \cite{cao-nature,cao-nature2,Yanko}. A theoretical analysis for the flatness was proposed in \cite{McD,tarno}.

  The model proposed in \cite{sanjose} consists of two Dirac Hamiltonians, each describing a layer, coupled by a periodic potential. The associated periodic lattice is hexagonal, with a larger pattern period than the one of the graphene sheets and dependent on the twisting angle. More precisely, we first rotate the primitive cell depicted in figure \ref{fig:ax} (with $\theta=\pi/3$) by an angle $-\pi/6$, and obtain the primitive vectors
$$
\ba_1=L \left(\frac{\sqrt{3}}{2},\frac{1}{2}\right), \qquad
\ba_2=L \left(\frac{\sqrt{3}}{2},-\frac{1}{2}\right), \qquad 
$$
where $L= a_0 \sqrt{1+3n+3n^2}$ for $n \in \NN$ when the twisting is commensurate and minimal. The parameter $a_0$ is the lattice constant of graphene and equal to $2.46$ Angstrom. The $x$ and $y$ axes are then aligned with the diagonals of the rhombus defined by $\ba_1$ and $\ba_2$. We nondimensionalize the latter to obtain unit vectors, and keep the same notations with an abuse. The primitive vectors of the reciprocal lattice are chosen to be
  $$
\bk_1=2 \pi \left(\frac{1}{\sqrt{3}},1\right), \qquad  \bk_2=2 \pi \left(-\frac{1}{\sqrt{3}},1\right).
$$
Following \cite{sanjose}, we introduce the operators
$$
\partial_\pm= -i \partial_{x_1}+\partial_{x_2} \mp (A_1 + i A_2), \qquad \textrm{with} \qquad \bA=(A_1,A_2)
= \left(0,\frac{ 2\pi}{3}\right).
$$
The coupling potentials $ V_{AA'}$, $V_{BA'}$ and $V_{AB'}$ defined in \cite{sanjose} are related by
\be \label{VAB}
V_{AA'}(\bx)=V(\bx), \qquad V_{BA'}(\bx)=V(\bx-\bv_0),\qquad V_{AB'}(\bx)=V(\bx+\bv_0),
\ee
where
$$
V(\bx)=t (1+e^{i \bk_1 \cdot \bx}+e^{i \bk_2 \cdot \bx}), \qquad \bv_0=(\ba_1+\ba_2)/3.
$$
Above, $t$ is the (nondimensional) strength of the coupling and is equal to $0.041 \sqrt{1+3n+3n^2}$. The Hamiltonian obtained in \cite{sanjose} is then
$$
H=
\begin{pmatrix}
0 & \partial_+ & V_{AA'} & V_{AB'}\\
\partial^*_+ & 0& V_{BA'} & V_{AA'}\\
V^*_{AA'} & V^*_{BA'} & 0 & \partial_-\\
V^*_{AB'} & V^*_{AA'} & \partial^*_- & 0 
\end{pmatrix},
$$
where $\partial^*_\pm$ is the adjoint of $\partial_\pm$.

We investigate now with our numerical method the stability of the flat bands under perturbations of the coupling potential $V$. We assume that the perturbation is random and consists of randomly located gaussians with random amplitudes, phases, and widths,  but still has the same period as the unperturbed structure and satisfies the relation in Eq.~\ref{VAB}. The latter assumption is not essential and may be violated for general disorder potentials, but is used here for computational convenience. More precisely, we set
$$
W(\bx)=\sum_{j=1}^{N_p} \alpha_j e^{i \theta_j} G\left( \frac{\bx-\bz_j}{\sigma_j}\right), \qquad G(\bx)=e^{-|\bx|^2}, 
$$
where the $\bz_j$ are uniformly drawn in the primitive cell, $\theta_j$, $\sigma_j$, $\alpha_j$ are uniformly distributed in $[0, 2 \pi]$, $[0.025,0.1]$, and $[0,\beta M_V]$ respectively, where $M_V$ is the maximal value of $|V|$, and $\beta$ a parameter that we will vary. The potential $V$ is then replaced by $V+W$ in the definition of the Hamiltonian. Note again that while the Fourier coefficients of $V$ are obvious, those of $W$ would have to be computed, hence limiting the use of the plane-wave expansion method. As in the last section, we set $N_1=N_2=25$. We represent for concreteness in figure \ref{fig:Pot} the absolute value of $V$ and of a realization of $W$ with $N_p=150$ and $\beta=0.15$. 

\begin{figure}[h!]
\centering
\includegraphics[height=5.5cm, width=7cm]{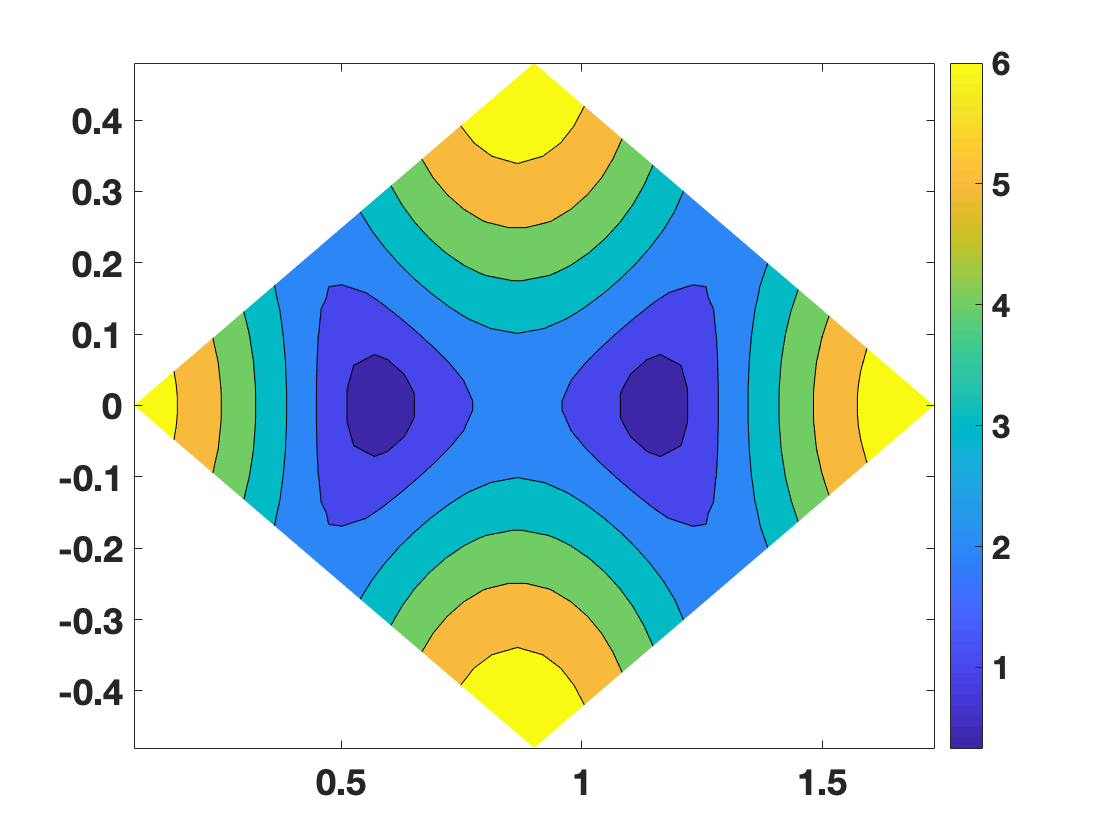}
    \includegraphics[height=5.5cm, width=7cm]{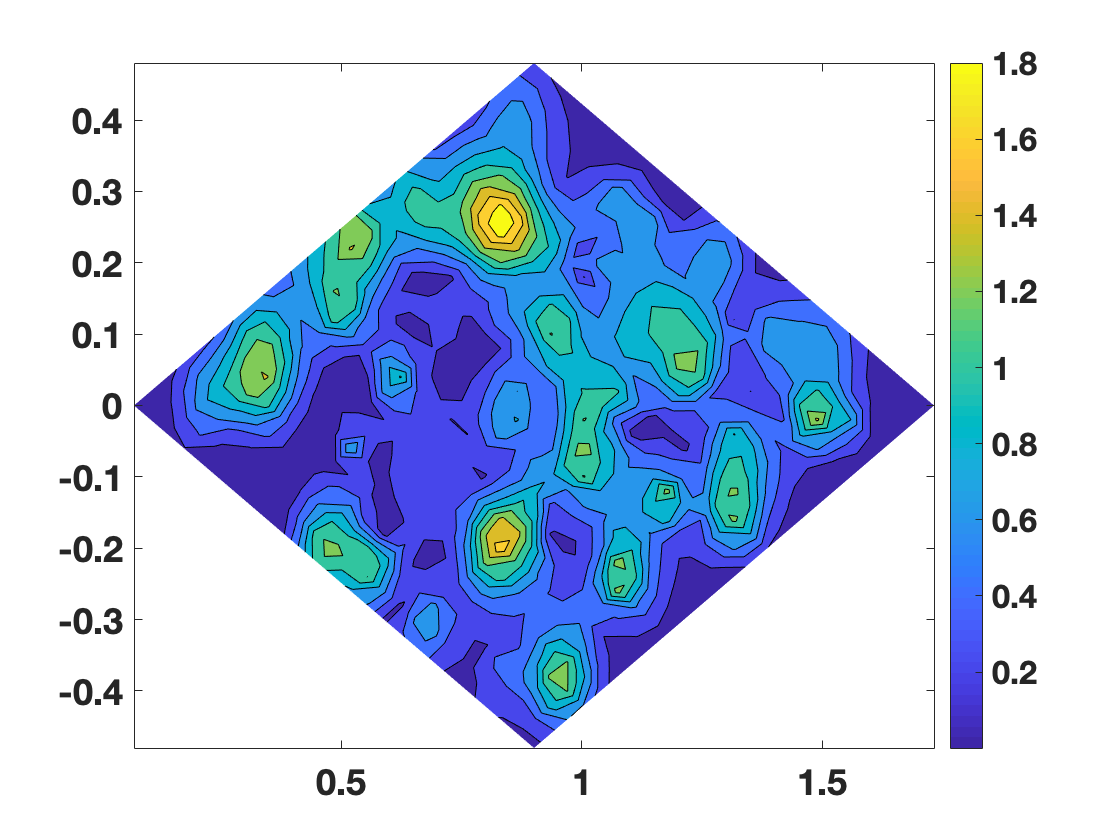} 

\caption{Left: $|V(\bx)|$. Right: $|W(\bx)|$ for $N_p=150$ and $\beta=0.015$.}
    \label{fig:Pot}
  \end{figure}

  We will depict the band structure by using the $\Gamma$, $K$ and $M$ points defined in the Brillouin zone that are frequently used in the physics literature. In our setting, they are defined with respect to the single layer Brillouin zone and take the values
  $$
  \Gamma=\left(-\frac{2 \pi}{\sqrt{3}},0\right), \qquad K=\left(0,\frac{2 \pi}{3}\right), \qquad M=\left(-\frac{\pi}{\sqrt{3}},\pi \right).
  $$
  The  $K$ point corresponds to one of the Dirac points. We move the point $\bk=(k_1,k_2)$ along straightlines connecting $\Gamma$, $K$ and $M$ to compute the band structure, and plot the bands along this path against $k_2=k_y$ in figure \ref{fig:bandt1}. In the left panel of figure \ref{fig:bandt1}, we plot the band structure for $\beta=0$, i.e. without random perturbations. For reference, we depict the Dirac cone at the point $K$ when there is no coupling (dotted lines in green), that is when $t=0$. The red dashed bands correspond to the case $n=20$, and the blue full lines to the flat band case with $n=35$. The choice $n=35$ leads to the smallest effective Fermi velocity at the $K$ point, with a magnitude more than 1000 times smaller than the Fermi velocity in each (uncoupled) layer.

  \begin{figure}[h!]
\centering
\includegraphics[height=5.5cm, width=7cm]{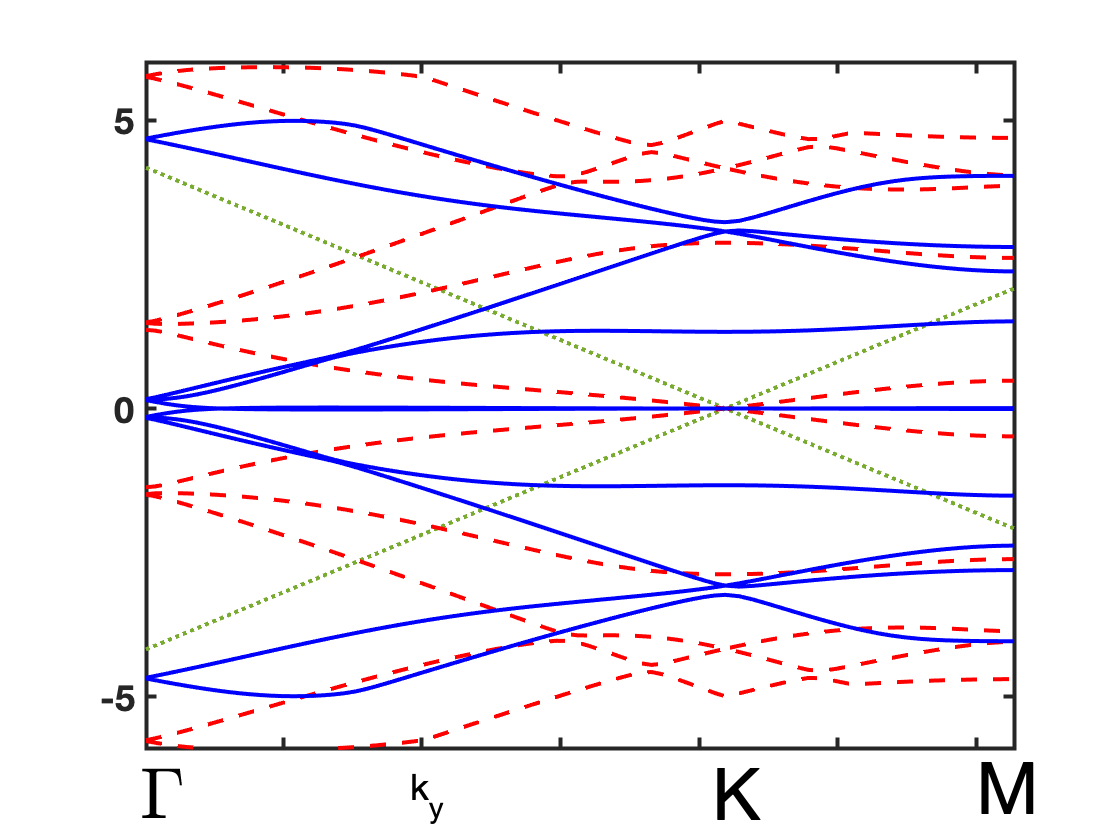}
\includegraphics[height=5.5cm, width=7cm]{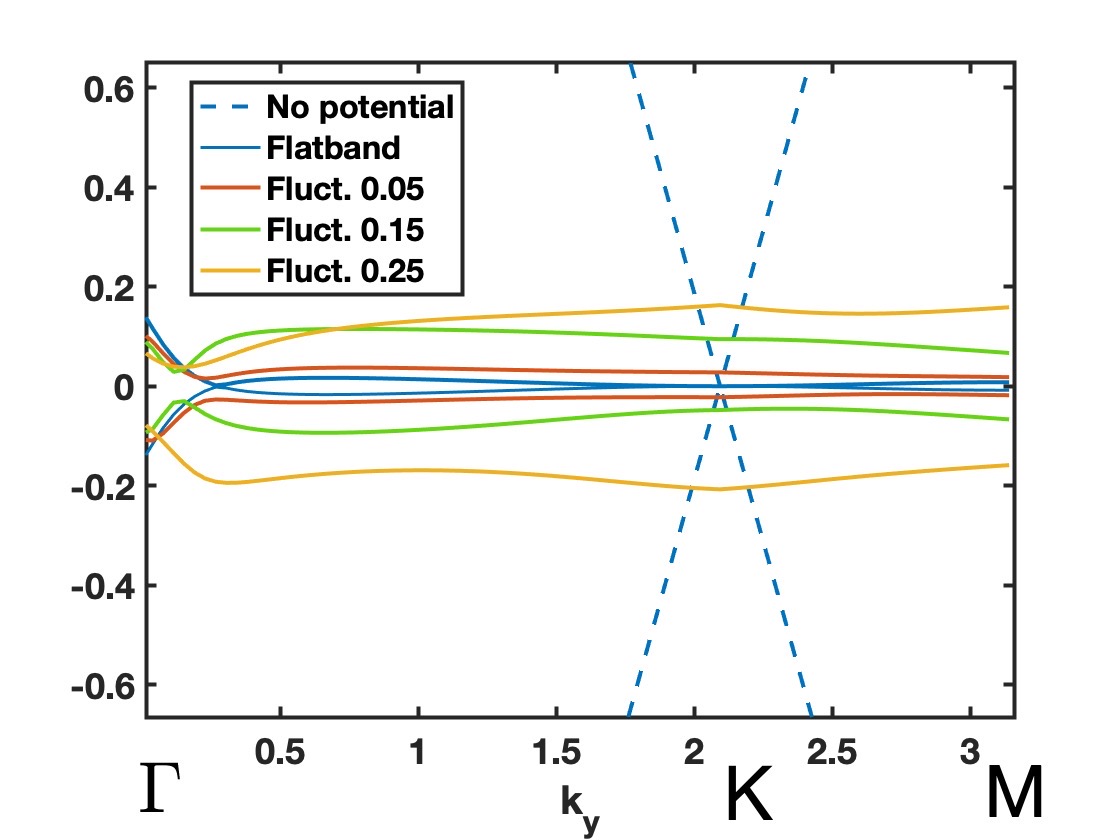}
\caption{Left panel: Band structure without random perturbations. The dotted lines correspond to the uncoupled case $t=0$, the red dashed bands correspond to the case $n=20$, and the blue full lines to the flat band case with $n=35$. Right panel: Perturbed band structure around the zero energy, for different fluctuations strength $\beta=0.05$ (red), $\beta=0.15$ (green), $\beta=0.25$ (yellow). The blue band corresponds to the unperturbed $\beta=0$ case.}
    \label{fig:bandt1}
\end{figure}

In the right panel of figure \ref{fig:bandt1}, we represent the bands around the zero energy in presence of perturbations of various magnitude of fluctuations, $\beta=0.05$ (red), $\beta=0.15$ (green), $\beta=0.25$ (yellow), along with the flat band and no coupling potential cases for comparison. One random realization for each amplitude is depicted. We zoom in around the Dirac point in the left panel of figure \ref{fig:bandt2} for a better assessment of the flatness. In the right panel of figure \ref{fig:bandt2}, we plot several random realizations in the case $\beta=0.05$ to evaluate the statistical stability of the bands. The blue line is the flat band and the red lines the random realizations.

The numerical results show that the random perturbations open a gap around the zero energy. While the zero-energy flat band is not stable, the random bands still remarkably exhibit flatness around the Dirac point, at least when the strength of the perturbations is not too large, say less than $\beta=0.25$. This is clearly observed in the left panel of figure \ref{fig:bandt2}. Flatness is stable in the sense that different realizations with similar strengths, here $\beta=0.05$, all exhibit flatness as shown in the right panel of figure \ref{fig:bandt2}. This stability suggests that the superconducting behavior observed in twisted bilayer graphene is at least robust under the particular perturbations considered here.
  
  \begin{figure}[h!]
\centering
\includegraphics[height=5.5cm, width=7cm]{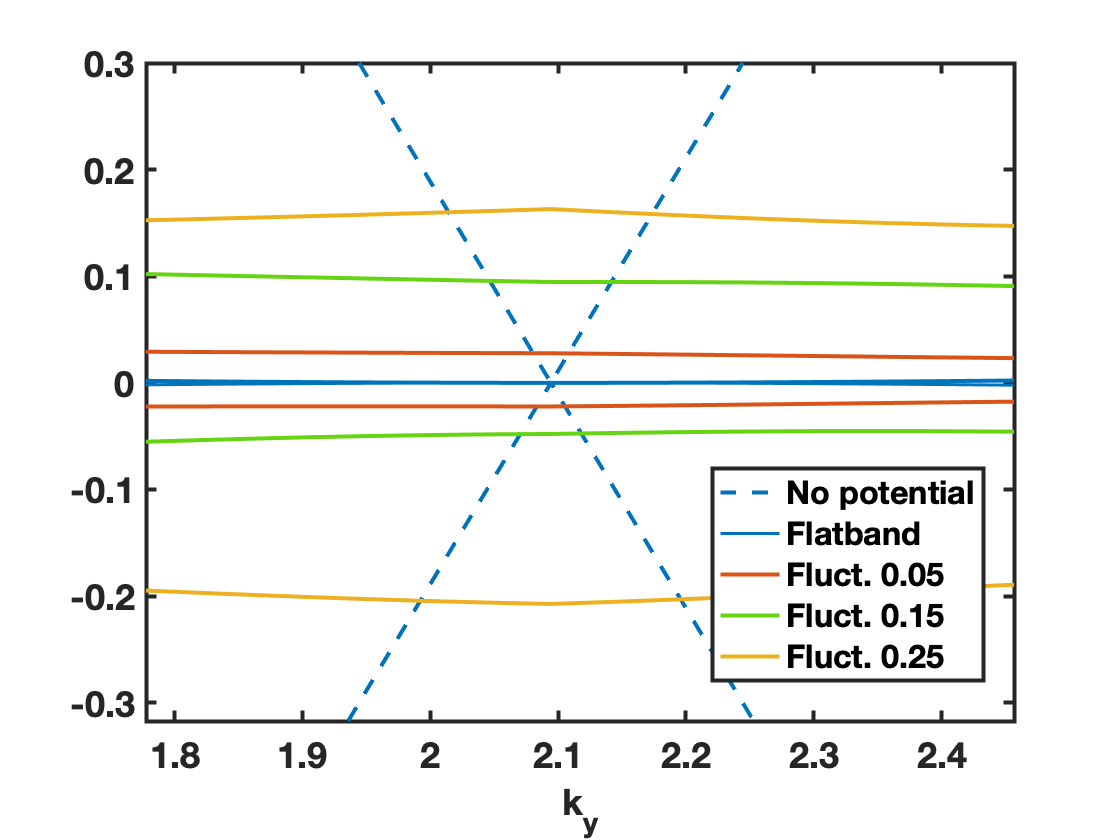}
    \includegraphics[height=5.5cm, width=7cm]{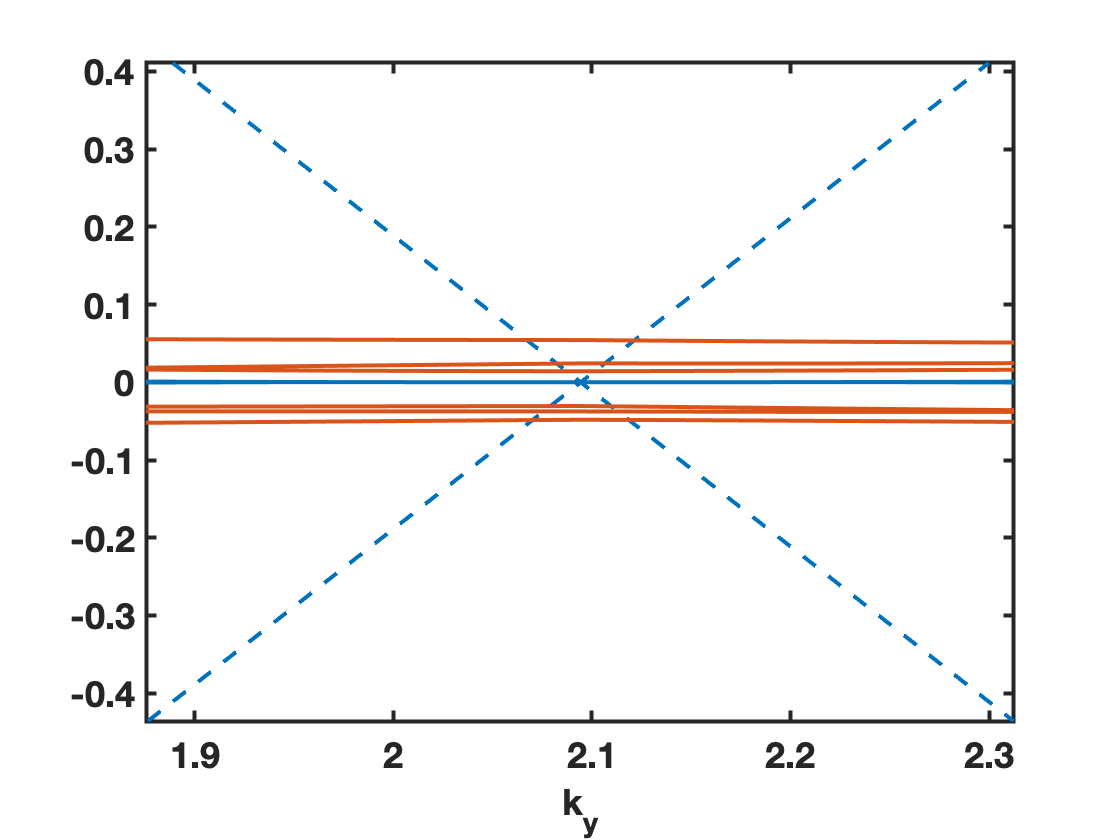} 
\caption{Left panel: right panel of figure \ref{fig:bandt1} zoomed in  around the Dirac point. Right panel: depiction of several random realizations of the bands (in red) for $\beta=0.015$. The blue band corresponds to the unperturbed $\beta=0$ case.}
    \label{fig:bandt2}
  \end{figure}
  
\section{Conclusion}
We derived in this work a numerical method for the diagonalization of periodic Dirac operators. The scheme is based on spectral methods that are immune to the Fermion doubling problem and provide us with high accuracy when the coefficients in the Hamiltonian are regular. The technique is applicable to all two-dimensional periodic lattices. An important point is to choose an odd number of nodes in the spatial discretization in order to obtain differentiation matrices with one-dimensional kernels spanned by constant vectors. We applied our scheme to the study of flat bands in graphene, and investigated in particular the stability of flat bands in twisted bilayer graphene under random perturbations. Our method can be directly generalized to three-dimensional periodic lattices.

\section{Appendix}
  \subsection{Proof of Lemma \ref{kernel}} Consider a vector $v \in \Rm^{N_1}$ with components $v_m$, $m=1,\cdots,N_1=2Q_1+1$, and extend $v$ by periodicity to negative indices, that is $v_{-m}=v_{N_1-m}$ for $0\leq m \leq Q_1$. Then, for $v \in N(T_1)$, the null space of $T_1$, we have 
    \be \label{eq1}
    (T_1 v)_j=\sum_{|m|\leq Q_1}v_{m} S'_{1}\big((j-m)h_1\big)=0, \qquad \textrm{for all} \quad j=-Q_1,\cdots,Q_1,
    \ee
    with by definition $(T_1 v)_{-j}=(T_1 v)_{N_1-j}$, for $j=0,\cdots,N_1$. Using periodicity, we rewrite \fref{eq1} as
    $$
    g_j:=(T_1 v)_j=\sum_{m=1}^{N_1} v_{m} S'_{1}\big((j-m)h_1\big)=0, \qquad j=1,\cdots,N_1,
    $$
    and define the discrete Fourier transform of a vector $(f_1,\cdots,f_{N_1})$ by
    $$
\hat f_k=\sum_{\ell=1}^{N_1} f_\ell e^{- 2 i \pi \ell k /N_1}, \qquad k=1,\cdots,N_1, \qquad \textrm{with} \quad \hat f_{k\pm N_1} =\hat f_k.
$$ 
Using the definition of $S_1$, we find, for $k=-Q_1,\cdots,Q_1$,
\bee\hat g_k&=&\sum_{\ell=1}^{N_1} \sum_{|m| \leq Q_1} \frac{2 i \pi m}{a_1} e^{2 i \pi \ell (m-k)/N_1} \left(\sum_{n=1}^{N_1} v_n e^{-2 i \pi  m n/N_1} \right)\\
&=&\sum_{|m| \leq Q_1}  \frac{2 i \pi m}{a_1} G(m-k) \hat v_m=0,
\eee
where
$$
G(n)=\sum_{\ell=1}^{N_1} e^{2 i \pi \ell n /N_1}= e^{2 i \pi n/N_1} \frac{1-e^{2 i \pi n}}{1-e^{2 i \pi n/N_1}}=N_1 \delta_{0,n}.
$$
Hence
$$
\hat g_k=N_1 \frac{2 i \pi k}{a_1} \hat v_k=0.
$$
As a consequence, $\hat v_k=0$ for $k=-Q_1,\cdots, Q_1$, with $k \neq 0$. Using the inverse formula
$$
v_\ell=\frac{1}{N_1}\sum_{|k| \leq Q_1} \hat v_k e^{2 i \pi k \ell /N_1},
$$
it follows that
$$
v_\ell=\frac{1}{N_1} \hat v_0, \qquad \ell=1, \cdots,N_1,
$$
which shows that $v_\ell$ is constant and concludes the proof.

\bibliographystyle{siam}
\bibliography{bibliography.bib}
\end{document}